\documentclass[12pt]{article}
\usepackage[T2A]{fontenc}
\usepackage[english]{babel}
\usepackage[english]{babel}
\usepackage[dvips]{graphicx}
\usepackage{mathtext}

\begin{document}

\title{Power spectrum of electron number density perturbations 
at cosmological recombination epoch}

\author{Venhlovska B., Novosyadlyj B.}
\maketitle

\medskip

\centerline{\it Astronomical Observatory of Ivan Franko National University of Lviv}
\medskip

\begin{abstract}

The power spectrum of number density perturbations of free electrons is obtained for the epoch of cosmological recombination of hydrogen. It is shown that amplitude of the electron perturbations power spectrum of scales larger than acoustic horizon exceeds  by factor of 17 the amplitude of baryon matter density ones (atoms and ions of hydrogen and helium). In the range of the first and second acoustic peaks such relation is 18, in the range of the third one  16. The dependence of such relations on cosmological parameters is analysed too.
\end{abstract}

\section*{Introduction}

The observational data on the cosmic microwave background (CMB) temperature fluctuations obtained in the ground-based, balloon and space experiments became the key test of cosmological models of the Universe and the main source for the determination of their parameters. The data obtained in the space mission WMAP (Wilkinson Microwave Anisotropy Probe) \cite{bennett2003,hinshaw2007,hinshaw2008} are the most outstanding achievement of the modern cosmology. They have given us the possibility to estimate the cosmological parameters (see \cite{spergel2007,apunevych2007,komatsu2008} and citing therein) with accuracy that practically equals to one of theoretical predictions because of the non completeness of physical processes governing the CMB temperature fluctuations and polarization  as well as the accuracy of analytical approximations and numerical computations. The important part of the theory of CMB anisotropy is the kinetics of cosmological recombination of primary plasma and calculation of free electrons number density at decoupling of thermal electromagnetic radiation and matter. The bases of the theory of cosmological recombination have been founded by Zeldovich \cite{zeldovich1968} and Peebles \cite{peebles1968} in 1968.  In the following papers \cite{matsuda1971, zabotin1982, lyubarsky1983, jones1985, krolik1990, rubicki1993} the main processes were particularly studied and a few-percent accuracy of calculations of recombination kinetics was achieved. The most complete analysis of the kinetics of cosmological recombination was carried out by Seager et al. (2000) \cite{seager2000}, where the multi-level ($\simeq300$ levels) structure of hydrogen and helium atoms, non-equilibrium  kinetics and, practically, all processes determining the thermal state of plasma were taken into account. The authors have created also the publicly available code  RECFAST \cite{seager1999}\footnote{the improved last release was done in September 2008}, which provides the computation of electron number density with accuracy $\sim 1\%$. It is used in the CMBFAST \cite{cmbfast96,cmbfast99}, CAMBCODE \cite{chalinor2005}, CMBEASY \cite{doran2005}  codes for calculation of power spectrum of matter density perturbations as well as CMB anisotropy. Since the accuracy of the WMAP observational data is close to the theoretical predictions one and is expected to be few times higher in the planned experiment PLANCK, the analysis of physical processes not taken into account until now  becomes actual. Among them are the transitions between the high levels of atoms of hydrogen and helium, their fine structure and elaboration of spontaneous, collision and induced transition rates, absorption and scattering of thermal radiation along the line of sight, isotope composition, molecular formation and dissociation, etc. In recent years these and other effects have been actively studied in order to improve an accuracy of computation of cosmological recombination \cite{leung2004,dubrovich2005,chluba2006a,kholupenko2006,chluba2006b}. In our papers \cite{novosyadlyj2006a,novosyadlyj2006b} the influence of adiabatic density perturbations on the number densities of ionized fractions has been studied. It was shown that at decoupling epoch the amplitude of relative perturbations of electron number density is 4-5 times higher than amplitude of relative perturbations of total baryon density. The reason for such difference is the considerably higher sensitivity of hydrogen and helium photoionization rate to temperature fluctuations than the spontaneous recombination rate one. It is clear, since the ionization of hydrogen and helium is provided by quanta of background thermal radiation of Wien spectrum range ($h\nu\gg kT$). The higher amplitude of relative density perturbations of free electron component, meanwhile, is not imprinted in the amplitude of angular power spectrum  of CMB temperature fluctuations in the linear approach\footnote{for second order effect see, for example, Khatri \& Wandelt, 2008, arXiv: 0810.4370 [astro-ph]}. However, it is not excluded that this effect has imprint in the map of the CMB primary polarization. Therefore, in this paper we study more detailed this effect for adiabatic perturbations of different scale at region of acoustic peaks to compare the power spectra of number density perturbations of different compoments at decoupling.
All computations were carried out using the code {\it drecfast.f} which is described particularly in \cite{novosyadlyj2006a,novosyadlyj2006b} and is publicly available at {\it http://astro.franko.lviv.ua/$\sim$novos/}. Research of the paper is restricted by the $\Lambda$CDM-model with parameters determined on the basis of data on CMB temperature anisotropy, large-scale structure and dynamics of expansion of the Universe. Combination of the WMAP data \cite{hinshaw2007} with different datasets on large-scale structure of the Universe, dynamics of its expansion, etc. gives somewhat different values of parameters. The ranges of values of cosmological parameters determined by \cite{spergel2007} are as follows:   $\Omega_{\Lambda}=0.7\div0.8$, $\Omega_m=0.23\div0.31$, $\Omega_{b}=0.04\div0.05$, ${\rm h}=0.68\div0.75$, $A_s=0.75\div0.92$, $n_s=0.9\div0.96$, where $\Omega_{\Lambda}\equiv\Lambda/3H_0^2$,  $\Omega_{b}\equiv\rho_b/\rho_{cr}$ $\Omega_m\equiv\rho_m/\rho_{cr}$ are cosmological constant, baryon matter density and total matter density (baryons + dark matter) in units of critical density $\rho_{cr}\equiv 3H_0^2/8\pi G$ respectively,
${\rm h}=H_0/100$km/s/Mpc is dimensionless Hubble constant, $A_s$ is amplitude of initial power spectrum of matter density perturbations, $n_s$ is spectral index of scalar mode of perturbations. Computations of cosmological recombination and power spectra of electron number density perturbations will be carried out for $\Lambda$CDM-model with two sets of the best-fit parameters:  $\Omega_{\Lambda}=0.736$, $\Omega_m=0.278$ $\Omega_{b}=0.05$, ${\rm h}=0.68$, $\sigma_8=0.73$, $n_s=0.96$ \cite{apunevych2007} and $\Omega_{\Lambda}=0.76$, $\Omega_m=0.24$ $\Omega_{b}=0.042$, ${\rm h}=0.73$, $A_s=0.83$, $n_s=0.958$ \cite{spergel2007}. The angular power spectra of CMB temperature fluctuations calculated for them are practically identical \cite{apunevych2007}.

\section{Dependence of relative number density of free electrons on redshit in the $\Lambda$CDM-model}

Important feature of the cosmological recombination of helium and hydrogen in the $\Lambda$CDM-model is that it occurs when the total energy density of thermal radiation approximately is comparable to baryon matter one: $\epsilon_{\gamma}\sim c^2\rho_b$. 
Henceforth we will use the following definitions:
$n_{\rm HI}$ and $n_{\rm HII}$ are number density of neutral and ionized hydrogen atoms;
$n_{\rm HeI}$, $n_{\rm HeII}$ and $n_{\rm HeIII}$ are number densities of neutral, singly and double ionized helium;
$n_e=n_{\rm HII}+n_{\rm HeII}+2n_{\rm HeIII}$ is number density of free electrons; 
$n_{\rm H}=n_{\rm HI}+n_{\rm HII}$ is total number density of hydrogen nuclei; $n_{\rm He}=n_{\rm HeI}+n_{\rm HeII}+n_{\rm HeIII}$ is total number density of helium nuclei. We use the relative number densities (ionization fractions): $x_{\rm HI}\equiv n_{\rm HI}/n_{\rm H}$ is relative abundance of neutral hydrogen, $x_{\rm HII}\equiv n_{\rm HII}/n_{\rm H}$ are relative abundances of ionized hydrogen,  $x_{\rm HeI}\equiv n_{\rm HeI}/n_{\rm He}$, $x_{\rm HeII}\equiv n_{\rm HeII}/n_{\rm He}$ and  $x_{\rm HeIII}\equiv n_{\rm HeIII}/n_{\rm He}$-- relative abundances of neutral, singly and double ionized helium, $x_e\equiv n_e/n_{\rm H}$ -- relative number density of electrons. The ratio of total number densities of helium and hydrogen nuclei we define as $f_{\rm He}\equiv n_{\rm He}/n_{\rm H}$, which can be expressed via mass fraction of primordial helium $Y_P$, so that $f_{\rm He}=Y_P/4(1-Y_P)$ (further we assume $Y_P=0.24$ \cite{schramm1998}). These quantities obey obvious relationships: $x_e=x_{\rm HII}+f_{\rm He}x_{\rm HeII}+2f_{\rm He}x_{\rm HeIII}$, $x_{\rm HI}+x_{\rm HII}=1$, $x_{\rm HeI}+x_{\rm HeII}+x_{\rm HeIII}=1$. The total mass density of baryons can be expressed via hydrogen number density and mass fraction of primordial helium in the following way: $\rho_{ b}\approx m_p n_{b}=m_p(n_{\rm H}+4n_{\rm He})=m_pn_{\rm H}(1+4f_{\rm He})$, where $n_{b}$ is mean number density of baryons (protons and neutrons), where $m_{ p}$ is mass of proton.

At early stages of evolution of the Universe ($z>10^4$) all hydrogen and helium atoms were ionized completely by thermal  photons, so $x_{\rm HII}=1$, $x_{\rm HI}=0$, $x_{\rm HeIII}=1$, $x_{\rm HeI}=x_{\rm HeII}=0$  and $x_e=1+2f_{\rm He}$ \cite{seager1999,seager2000}.  At $z\sim 8000$ thermal photons with energies higher than ionization potential 
of ${\rm HeII}$ from ground level reside in the short-wave tail of Planck function and their number density becomes too low to keep all helium in the ionization state of ${\rm HeIII}$. It begins to recombine and $\rm HeII$ ions  appear. Recombination of $\rm HeII$ occurs in the conditions of local thermodynamic equilibrium (LTE), so, the ionization fraction of helium, $x_{\rm HeIII}$, is described by Saha equation:
\begin{equation}
  \label{SahaHeIII}
  {x_ex_{\rm HeIII}\over x_{\rm HeII}} =
  {(2\pi m_e k T_m)^{3/2}\over h^3 n_{\rm H}}
  e^{-\chi_{\rm HeII}/kT_m},
\end{equation}
where $T_m$ is matter temperature, $m_{e}$ is mass of electron, $h$ is Planck constant, $k$ is Boltzmann constant, $\chi _{\rm HeII}$ is ionization potential of HeII.
Since at this epoch both hydrogen and helium are completely ionized  ($x_{\rm HI}=0$, $x_{\rm HII}=1$, $x_{\rm HeI}=0$) we have $x_{\rm HeII}=1-x_{\rm HeIII}$ and
$x_e=1+f_{\rm He}(1+x_{\rm HeIII})$, so equation (\ref{SahaHeIII}) can be easily solved for $x_e$. Using it one can easily check that  already at 
$z\sim 5000$ all helium atoms become singly ionized. Such state is kept up to $z\sim 3500$ when $\rm HeI$ begins to recombine. 
At this time the conditions are close to LTE yet.  
The metastable $2s$ level plays insignificant role in deviation 
of radiative recombination rate of ${\rm HeI}$ from LTE one until the part of HeI is less than 1\% of  total  helium content and ionized fraction $x_{\rm HeII}$ is described yet enough accurately by Saha equation: 
\begin{equation}
  \label{SahaHeII}
  {x_ex_{\rm HeII}\over x_{\rm HeI}} =
  4{(2\pi m_e k T_m)^{3/2}\over h^3 n_{\rm H}}
  e^{-\chi_{\rm He I}/kT_m},
\end{equation}
where $\chi _{\rm HeI}$ is ionization potential of HeI.
Now $x_{\rm HeIII}=0$ and $x_{\rm HeI}= 1-x_{\rm HeII}$. For accurate calculation of $x_{\rm HeII}$ we must have the exact value of $x_e=x_{\rm HII}+f_{\rm He}x_{\rm HeII}$. Despite $x_{\rm HII}\approx1$, the decrease of $n_{\rm HII}$ in 0.1\% caused by the hydrogen recombination leads to the variation of $n_e$ comparable to one caused by ${\rm HeI}$ recombination because of the domination of hydrogen ($f_{\rm H}=n_{\rm H}/(n_{\rm H}+n_{\rm He})=0.921$). So, at this step the hydrogen recombination already must be taken into account too. The ionized fraction $x_{\rm HII}$ is described yet enough accurately by Saha equation:
\begin{equation}
  \label{SahaHII}
  {x_ex_{\rm HII}\over x_{\rm HI}} =
  {(2\pi m_e k T_m)^{3/2}\over h^3 n_{\rm H}}
  e^{-\chi_{\rm HI}/kT_m},
\end{equation}
where $\chi _{\rm HI}$ is ionization potential of HI.
The system of these two equations can be reduced to the single cubic equation for $x_e$, which has one real root:
\begin{equation}
  \label{xe3000}
  x_e=2\sqrt{-A/3}\cos{(\alpha/3)}-B/3,
\end{equation}
where $B=R_{\rm HI}+R_{\rm HeI}$, $R_{\rm HeI}$ and $R_{\rm HI}$ is right hand of equations (\ref{SahaHeII}) and (\ref{SahaHII}),
$\cos{\alpha}=C/2\sqrt{-A^3/27}$, $A=D-B^2/3$, $D=R_{\rm HI}R_{\rm
  HeI}-R_{\rm HI}-f_{\rm He}R_{\rm HeI}$, $C=2B^3/27-BD/3-E$,
$E=-R_{\rm HI}R_{\rm HeI}(1-f_{\rm He})$. The code RECFAST was complemented by this solution in order to achieve more accurate computations of number density perturbations of ions. However, it does not change the results of calculations of unperturbed $x$'s noticeably \cite{novosyadlyj2006a,novosyadlyj2006b}.

Metastable levels 2s HeI and HI cause a delay of recombination of HeII$\rightarrow$HeI and HII$\rightarrow$HI, violation of LTE population of levels and equilibrium of ionization-recombination processes (''bottleneck'' effect). Saha equation does not already describe adequately the recombination 
and equations of detailed balance must be used \cite{seager1999}:
\begin{eqnarray}
  \label{eqHeII}
  {dx_{\rm HeII}\over dz} =
  \left(x_{\rm HeII}x_e n_{\rm H} \alpha_{\rm HeI}
    - \beta_{\rm HeI} (1-x_{\rm HeII})
    e^{-h\nu_{\rm HeI2^1s}/kT_m}\right)\nonumber \\
  \times {1 + K_{\rm HeI} \Lambda_{\rm He} n_{\rm H}
    (1-x_{\rm HeII})e^{-h\nu_{ps}/kT_m}
    \over H(z)(1+z)\left(1+K_{\rm HeI}
      (\Lambda_{\rm He} + \beta_{\rm HeI}) n_{\rm H} (1-x_{\rm HeII})
      e^{-h\nu_{ps}/kT_m}\right)},
\end{eqnarray}
and
\begin{eqnarray}
  \label{eqHII}
  {dx_{\rm HII}\over dz} = \left(x_ex_{\rm HII} n_{\rm H} \alpha_{\rm H}
    - \beta_{\rm H} (1-x_{\rm HII})
    e^{-h\nu_{\rm HI 2s}/kT_m}\right)\nonumber \\
  \times{1 + K_{\rm H} \Lambda_{\rm H} n_{\rm H}(1-x_{\rm HII})
    \over H(z)(1+z)\left(1+K_{\rm H} (\Lambda_{\rm H} + \beta_{\rm H})
      n_{\rm H} (1-x_{\rm HII}) \right)},
\end{eqnarray}
where
\begin{eqnarray}
  \label{alphaHeI}
  \alpha_{\rm HeI} =q\left[\sqrt{T_m\over T_2}\left(1+\sqrt{T_m\over T_2}\right)^{1-p}
    \left(1+\sqrt{T_m\over T_1}\right)^{1+p}\right]^{-1}\!   \mathrm {m^{3}c^{-1}},
\end{eqnarray}
\begin{eqnarray}
  \label{alphaHI}
  \alpha_{\rm H} = F\cdot 10^{-19}at^{b}/(1 + ct^{d}),\, \mathrm{m^{3}c^{-1}} \;\;\;  t=T_m/10^4.
\end{eqnarray}
$\alpha_{\rm HeI}$ and $\alpha_{\rm H}$ are effective recombination coefficients of helium \cite{hummer1998} and hydrogen \cite{pequignot1991}, respectively.
$K_{\rm HeI}\equiv \lambda^3_{\rm HeI2^1p}/[8\pi H(z)]$ is the factor taking into account the cosmological redshifting of HeI  $2^1p-1^1s$  photons and $K_{\rm H}\equiv \lambda^3_{\rm H2p}/[8\pi H(z)]$ is the factor taking into account the cosmological redshifting of
Ly$\alpha$ photons. The effective photoionization coefficients in  (\ref{eqHeII}) and  (\ref{eqHII}) are calculated via effective recombination coefficients as follows:
\begin{eqnarray}
  \label{beta}
  \beta=\alpha (2\pi m_e k T_m/h^2)^{3/2}e^{-h\nu_{2s-1s}/kT_m}.
\end{eqnarray}

Before $z\sim 800$ the matter temperature $T_m$ practically equals radiation temperature $T_{\rm R}$ since until the time-scale of Thomson scattering remains essentially lower than the time-scale of expansion of the Universe, $t_{\rm T}/t_{\rm Hubble}<10^{-3}$. Therefore, the rate of temperature decreasing is governed by adiabatic cooling of radiation ($\gamma=4/3$) caused by expansion of the Universe:
\begin{equation}
  \label{Tm1}
  \frac{dT_m}{dz} =\frac{T_m}{(1+z)}.
\end{equation}
And only after recombination at $z<800$ adiabatic cooling of ideal gas ($\gamma=5/3$) begins to dominate over the heating caused by Compton effect which is the main process of energy transfer between electrons and photons. 
Cooling of plasma via free-free, free-bound and bound-bound transitions and collisional ionization as well as heating via photoionization and collisional recombination gives insignificant contribution into the rate of temperature change, it does not  exceed the  0.01$\%$ of main processes -- adiabatic cooling and heating by Compton effect  \cite{seager2000}. So, at this epoch the following equation for rate of temperature decreasing is enough accurate \cite{seager1999}:
\begin{equation}
  \label{Tm2}
  \frac{dT_m}{dz} = \frac{8\sigma_{\rm Th}a_{\rm R}
    T_{\rm R}^4}{3H(z)(1+z)m_ec}\,
  \frac{x_e}{1+f_{\rm He}+x_e}\,(T_m - T_{\rm R})
  + \frac{2T_m}{(1+z)}, \nonumber
\end{equation}
The Table\ref{tab} lists the values of all atomic constants and coefficients used in the equations (\ref{SahaHeIII})-(\ref{Tm2}).

\begin{figure}
\centerline{\includegraphics[height=10cm]{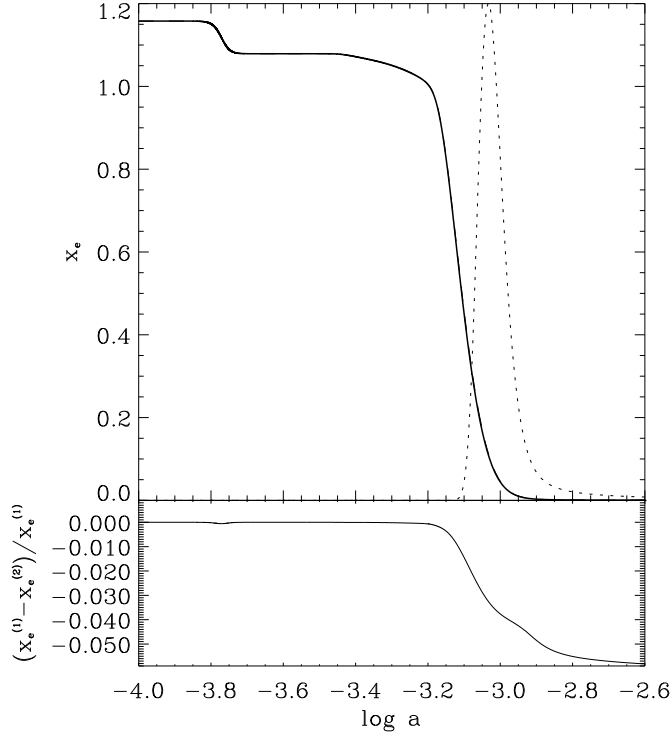}}
\caption{Top panel: the dependence of relative number density of free electrons on redshift in $\Lambda$CDM-model with parameters $\Omega_{\Lambda}=0.736$, $\Omega_m=0.278$, $\Omega_{b}=0.05$, ${\rm h}=0.68$ \cite{apunevych2007} (solid line) and  $\Omega_{\Lambda}=0.76$, $\Omega_m=0.24$, $\Omega_{b}=0.042$, ${\rm h}=0.73$ \cite{spergel2007} (dashed line overlaps with solid one). Dotted line represents visibility function $d\tau /dz e^{-\tau}$ ($\times 270$). The relative difference $x_e$ for two sets of parameters of $\Lambda$CDM-model is shown in bottom panel.}
\label{xe}
\end{figure}

In Fig.\ref{xe} the relative number density of electrons $x_{e}$ at range $400\le z\le 10000$ is presented.
The visibility function $d\tau /dz e^{-\tau}$ (dotted line) shows the region of the largest cosmological recombination rate and decoupling epoch -- here $\tau$ is optical depth caused by Thomson scattering by electrons, $z=(a^{-1}-1)$ is redshift.

The calculations shown in Fig.\ref{xe} have been done for $\Lambda$CDM model with parameters  $\Omega_{\Lambda}=0.736$, $\Omega_m=0.278$, $\Omega_{b}=0.05$, ${\rm h}=0.68$ \cite{apunevych2007} and $\Omega_{\Lambda}=0.76$, $\Omega_m=0.24$, $\Omega_{b}=0.042$, ${\rm h}=0.73$ \cite{spergel2007}. The curves  $x_e(a)$ for two sets of parameters practically overlap: the difference $\sim4\%$ is in the range of maximum of visibility function ($z_{dec}\approx1080$) and increases up to $\sim6\%$ for residual ionization $x_e\sim 10^{-3}$ at $z\le 900$.

\section{Perturbations of number density of ions and electrons }

\subsection{Definitions}

Let the values of ionization fractions of HI, HII, HeI, HeII, HeIII and electrons averaged over the whole space at fixed cosmological time be $x_i$, where {\it ''i''} marks each component. Let us denote the local value of relative number density of each component in the range of cosmological density perturbation of baryon matter $\delta_{b}\equiv \delta \rho_b/ \rho_b\ll1$, were $\rho_b$ is mean matter density of baryons, as $\hat x_i$. Its deviation from mean 
value we denote by $\delta x_i$, so that $\hat x_i= x_i + \delta x_i$ and $\delta x_i$ is called the perturbation of relative number density (fraction) of the {\it i}-th component. We define relative perturbations of ions and free electrons fractions as 
$\Delta_i\equiv \delta x_i/ x_i$, where $x_i$ is any of them. It is obvious that
$\Delta_e=\delta n_{e}/ n_{e} - \delta n_{\rm H}/ n_{\rm H}$, 
$\Delta_{\rm HII}=\delta n_{\rm HII}/ n_{\rm HII} - \delta n_{\rm H}/ n_{\rm H}$, 
$\Delta_{\rm HeII}=\delta n_{\rm HeII}/ n_{\rm HeII} - \delta n_{\rm He}/ n_{\rm He}$, 
$\Delta_{\rm HeIII}=\delta n_{\rm HeIII}/ n_{\rm HeIII} - \delta n_{\rm He}/ n_{\rm He}$.
We suppose the primordial chemical composition of baryon matter to be uniform ($f_{\rm He}$ is constant) and for homogeneous medium\footnote{It is also macroscopically electroneutral: $n_e=n_p+n_{\rm HeII}+2n_{\rm HeIII}$ everywhere and always.} 
$\delta n_{\rm H}/ n_{\rm H}=\delta n_{\rm He}/ n_{\rm He}=\delta_{b}$, then 
\begin{equation}
\label{deltax}
\Delta_i=\delta_i -\delta_{b}, 
\end{equation}
where $\delta_i\equiv\delta n_{i}/ n_{i}=\Delta_i+\delta_{b}$ is relative number density perturbation of {\it i}-th component. It must be noted
that in expanding Universe the recombination does not end with the completely neutral hydrogen or helium but with residual ionization. Therefore, none of values $n_i$ reach zero and ambiguity of $"0/0"$-type in $\delta_i$ does not appear.
So, $\Delta_i$'s never diverge. Numerical results presented in \cite{novosyadlyj2006a,novosyadlyj2006b} and below prove that.

Therefore, $\Delta_i$ is difference of number densities relative perturbations of {\it ''i''}-th component and of all baryon matter. Since $\delta_i$ and  $\delta_{b}$ are scalar functions of four coordinates in some gauge, under the gauge transformations not changing the cosmological background each of them is transformed by adding the same expression from the time coordinate transformation component (see, for example, \cite{bardeen1980,kodama1984,durrer2001a}). As soon as they appear in (\ref{deltax}) with opposite signs, $\Delta_i$'s keep unchanged under such transformations, so they are gauge-invariant variables. 

If hydrogen and helium are entirely ionized and ionization degree does not change with time then 
$\delta_i=\delta_{b}$ and $\Delta_i=0$. If the photorecombination and photoionization rates as well as ionization degree of some
component change in space and time then  $\delta_i$ and $\delta_{b}$ can evolve with different rates because the variation of $\delta_{b}$ is driven by gravitation and stress of baryon-photon plasma and $\delta_i$ is additionally influenced by kinetics of ionization-recombination processes. Therefore,
$\Delta_i$ is a measure of deviation of relative number density perturbation of {\it ''i''}-th component from relative density perturbation of total baryon component $\delta_{b}$, caused by different recombination and ionization rates within region of cosmological density perturbation.

At early stage of the Universe evolution the adiabatic relative density perturbations of baryon matter $\delta_{b}$ and radiation energy $\delta_{\rm R}\equiv \delta \epsilon_{\rm R}/
\epsilon_{\rm R}$ obey the following relation: $\epsilon_{\rm R}=4\delta_{b}/3$. Since $\epsilon_{\rm R}=aT_{\rm R}^4$, $\delta_{T_R}\equiv \delta T_R/ T_R=1/3\delta_b$.

\subsection{Equations}

The amplitudes of the baryon density $\delta_b$ and thermal radiation $\delta_{T_R}$ perturbations generated in the early Universe increase because of gravitational instability and at the moment of recombination they achieve a value $\simeq 10^{-4}-10^{-5}$ at scales 30-300$h^{-1}$Mpc (they also depend on power spectrum of initial perturbations). The local baryon mass density perturbations lead most probably to corresponding perturbations of number density of ions and electrons, $\delta_e \propto \delta_b$. Since the rates of ionization-recombination processes depend on density and temperature of baryon matter and radiation, in the region of the perturbations the departure of the distribution of atoms over ionization states from the background one will occur, so $\Delta_i\ne 0$ is expected. We study the cosmological perturbations of small amplitudes. It means that within region of cosmological perturbations all equations (\ref{SahaHeIII})-(\ref{Tm2}) are applicable and connection between the perturbations of ion number density and cosmological perturbations of density and temperature can be obtained by variation of those equations.

Varying the equation (\ref{SahaHeIII}) for electron ionization fraction $\Delta_{e}= x_{\rm HeIII}f_{\rm He}\Delta_{\rm HeIII}/x_{e}$  we got:
\begin{eqnarray}
  \label{SahadxHeIII}
  \Delta_{e} &=& {x_{\rm HeIII}(1-x_{\rm HeIII})f_{\rm He}\over x_e+(1-x_{\rm HeIII})x_{\rm HeIII}
f_{\rm He}}\left[\left({3\over 2}+{\chi_{\rm HeII}\over kT_m}\right)
      \delta_{T_m}-\delta_b\right].
\end{eqnarray}
We see that relative perturbation of electron number density is linear combination of initial relative perturbations of baryon matter temperature and density.
In the region of adiabatic perturbations the fluctuations $\Delta_e$ and $\delta_{T_m}$ have the same sign and opposite one to the baryon density perturbation $\delta_b$. The values of $x_e$ and $x_{\rm HeIII}$ are calculated from (\ref{SahaHeIII}). The asymptotical behaviour of $\Delta_e$ follows from (\ref{SahadxHeIII}): at $z>7000$ when $x_{\rm HeIII}\to 1$ (all helium atoms become double ionized) $\Delta_{{\rm HeIII}}\to 0$ ($\delta_{n_{\rm
    HeIII}}=\delta_{b}$) and at redshift $z<5000$ when $x_{\rm HeIII}\to 0$ (all helium atoms become singly ionized) $\Delta_e\to 0$. So, $\Delta_e$ has peak in this range of redshifts (see Fig.2 in \cite{novosyadlyj2006b}).

At $3500<z<5000$ both hydrogen and helium are entirely ionized (helium singly): $x_{\rm HII}=x_{\rm HeII}=1$, $x_{\rm HI}=x_{\rm HeI}=x_{\rm HeIII}=0$, so the amplitudes of all relative perturbations equal to zero. With subsequent decreasing of temperature HeI atoms and afterwards HI ones begin to recombine. The kinetics of their recombination is described by Saha equations  (\ref{SahaHeII}) and (\ref{SahaHII}). Variation of these equations gives the expressions for relative perturbations of helium 
$\Delta_{\rm HeII}$ and hydrogen $\Delta_{\rm HII}$ fractions, using them the relative perturbation of free electrons fraction $\Delta_e$ can be presented in the form: 
\begin{eqnarray}
  \label{Sahadxe}
  \Delta_{e} = {(1-x_{\rm HII})x_{\rm HII}{\chi_{\rm HI}\over kT_m}\left(1+(1-x_{\rm HeII})x_{\rm HeII}/x_e\right)\over
    (1-x_{\rm HII})x_{\rm HII}+(1-x_{\rm HeII})x_{\rm HeII}f_{\rm He}-x_e}\delta_{T_m}+  \nonumber \\
  +{(1-x_{\rm HeII})x_{\rm HeII}{\chi_{\rm HeI}\over kT_m}\left(f_{\rm He}-(1-x_{\rm HII})x_{\rm HII}/x_e\right)\over
    (1-x_{\rm HII})x_{\rm HII}+(1-x_{\rm HeII})x_{\rm HeII}f_{\rm He}-x_e}\delta_{T_m}+  \nonumber \\
  + {(1-x_{\rm HII})x_{\rm HII}+(1-x_{\rm HeII})x_{\rm HeII}f_{\rm He}\over 
    (1-x_{\rm HII})x_{\rm HII}+(1-x_{\rm HeII})x_{\rm HeII}f_{\rm He}-x_e}\left[{3\over 2} \delta_{T_m}-\delta_b\right].
\end{eqnarray}
At $x_{\rm HeII}\to 1$ and $x_{\rm  HII}\to 1$  $\Delta_{e}$ $\to
0$ as expected. Another asymptotical behaviour ($x_{\rm HeII}\to 0$ and $x_{\rm HII}\to 0$) has not physical sense as soon as
at $x_{\rm HeII}\le 0.99$ and $x_{\rm HII}\le 0.99$ it is necessary to use the non-equilibrium kinetics equations and energy balance (\ref{eqHeII})-(\ref{Tm2}). In this case the differential equations for relative perturbations $\Delta_{\rm HII}$, $\Delta_{\rm HeII}$  and $\delta_{T_m}$ can be obtained by the variation of (\ref{eqHeII})-(\ref{Tm2}). Such equations in the explicit form are presented in  \cite{novosyadlyj2006a,novosyadlyj2006b}, their generalized form is following:
\begin{eqnarray}{}
  \label{eqdxi}
  &x_{i}\frac{d \Delta_{i}}{dz} =\frac{dx_{i}}{dz}\left[A_{i}\Delta_{e}+\left\lbrace \frac{x_{i}}{1-x_{i}}\left(B_{i}-C_{i}+D_{i} \right)+A_{i}-1 \right\rbrace \Delta_{i}+\left\lbrace A_{i}+C_{i}-D_{i} \right\rbrace\delta_{b}+\right. & \nonumber \\
  &\left. \left\lbrace F_{i}\left( 1-D_{i}\frac{\beta_{i}}{\Lambda_{i}+\beta_{i}}\right)- \left(B_{i}+D_{i}\frac{\beta_{i}}{\Lambda_{i}+\beta_{i}}\right) \left( \frac{3}{2}+\frac{h\nu_{2si}}{kT_{m}}\right)-B_{i}\frac{h\nu_{i2s}}{kT_{m}}+ \Theta_{i}\frac{h\nu_{ps}}{kT_{m}}\left(C_{i}-D_{i}\right)\right\rbrace \delta_{T_{m}} \right].
  \end{eqnarray}
where index of ''i'' has two values corresponding to HeII or HII and coefficients $A_{i}$, $B_{i}$, $C_{i}$, $D_{i}$, $F_{i}$, $\Theta_{i}$ are determined by number density of ions of helium or hydrogen and the recombination rates (see Appendix). The equations of non-equilibrium recombination (\ref{eqHeII}) and (\ref{eqHII}) are written as follows:
\begin{eqnarray}{}
 \label{eqxi}
 \frac{dx_{i}}{dz} = \frac{x_{i}x_{e}n_{\rm H}\alpha_{i}}{H\left(z \right)\left(1+z
\right)}\frac{D_{i}}{A_{i}C_{i}}\frac{\Lambda_{i}}{\Lambda_{i}+\beta_{i}}.
\end{eqnarray} 
The expression for relative perturbations of free electrons fraction can be found from (\ref{eqdxi})--(\ref{eqxi}):  
\begin{eqnarray}{}
  \label{eqdxe}
  &\Delta_{e} = \frac{H \left(z \right) \left(1+z \right)}{x_{e}n_{H} \alpha_{i}}\frac{\Lambda_{i}+\beta_{i}}{\Lambda_{i}}\frac{C_{i}}{D_{i}}\frac{d\Delta_{i}}{dz}-
  \frac{1}{A_{i}}\left[\left\lbrace \frac{x_{i}}{1-x_{i}}\left(B_{i}-C_{i}+D_{i} \right)+A_{i}-1 \right\rbrace \Delta_{i}+\left\lbrace A_{i}+C_{i}-D_{i} \right\rbrace\delta_{b}+\right. & \nonumber \\
  &\left. \left\lbrace F_{i}\left( 1-D_{i}\frac{\beta_{i}}{\Lambda_{i}+\beta_{i}}\right)- \left(B_{i}+D_{i}\frac{\beta_{i}}{\Lambda_{i}+\beta_{i}}\right) \left( \frac{3}{2}+\frac{h\nu_{2si}}{kT_{m}}\right)-B_{i}\frac{h\nu_{i2s}}{kT_{m}}+ \Theta_{i}\frac{h\nu_{ps}}{kT_{m}}\left(C_{i}-D_{i}\right)\right\rbrace \delta_{T_{m}} \right].
\end{eqnarray} 
The evolution of relative perturbations of baryon matter temperature is described by equation:
\begin{eqnarray}{}
  \label{eqdTm}
  &T_m\frac{d\delta_{T_m}}{dz} = \frac{8\sigma_{\rm Th}a_{\rm R}
    T_{\rm R}^4}{3H(z)(1+z)m_ec} 
  \frac{x_e}{1+f_{\rm He}+x_e} \left[\left(T_m-T_{\rm R} \right)\frac{1+f_{\rm He}}{1+f_{\rm He}+x_e}\Delta_e+\left(4T_m-5T_{\rm R} \right)\delta_{T_{R}} +T_{\rm R}\delta_{T_{m}} \right].&
\end{eqnarray}

Thus, the system of first-order ordinary linear differential equations for relative perturbations of matter temperature and ions and electrons fractions consists of equations (\ref{eqdxi})-(\ref{eqdTm}) and can be solved using the publicly available code DVERK \cite{dverk}.

The equations (\ref{SahadxHeIII})-(\ref{eqdTm}) were used to analyse the evolution of relative density perturbations of ions and temperature perturbations of baryonic matter. All equations contain the solutions of unperturbed problem therefore it seems naturally to supplement the code RECFAST \cite{seager1999} with block calculating the perturbations of ionization fractions. The complemented code $drecfast.f$ \cite{drecfast} is used further in our analysis of evolution of number density perturbations of free electrons.

\section{Evolution of relative density perturbation of free electrons}

Before recombination the time variations of baryons density and thermal radiation temperature perturbations depend on the relation of scale of perturbations to acoustic horizon scale \cite{lifshitz1946}. 
When the scale of perturbation becomes substantially smaller than scale of acoustic horizon (Jeans scale) before recombination, then  adiabatic perturbations in the baryon-photon plasma start to oscillate like the standing acoustic waves. In consequence of recombination the Jeans scale drops and the previously oscillating amplitudes of perturbations in baryon component start to increase monotonously mainly as a result of gravitational attraction of dark matter density perturbations. The 
amplitudes of perturbations with scales larger than acoustic horizon at recombination epoch increased as $\delta_b\propto t^{1/2}$ in radiation-dominated epoch and $\delta_b\propto t^{2/3}$ after recombination in dust-like Universe. In  the papers  \cite{bardeen1980,kodama1984,ma1995,hu1995a,durrer2001a,novosyadlyj2007} one can find the analytical solutions of relevant equations for evolution of relative density perturbations in simplified cases of single component media as well as the numerical solutions for real multi-component Universe. 

\begin{figure}
\centerline{\includegraphics[height=16cm]{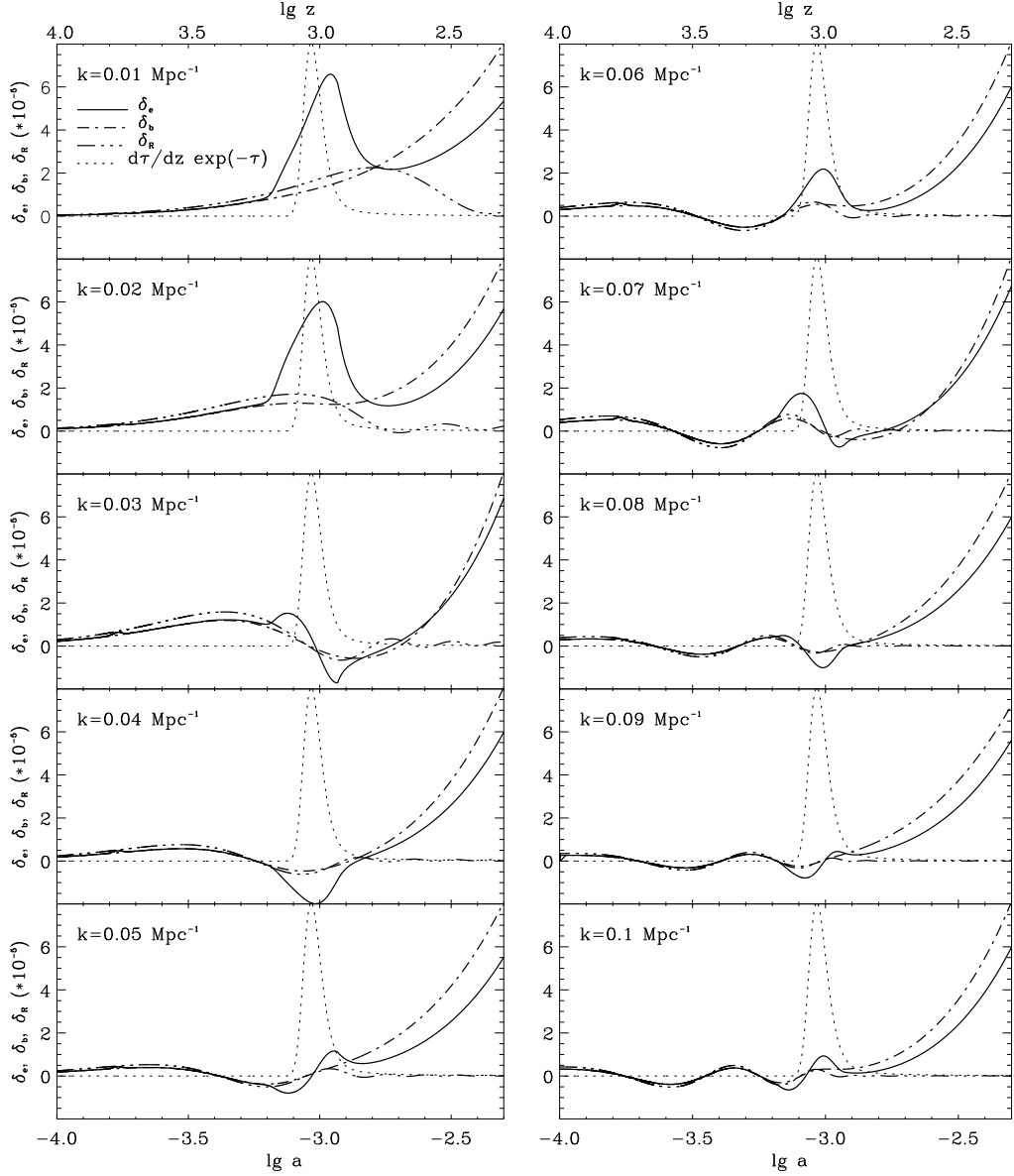}}
\caption{The evolution of number density perturbations of free electrons in region of adiabatic matter density perturbations in $\Lambda{\rm CDM}$-model \cite{apunevych2007} for ${\rm k}=0.01,...0.1$Mpc$^{-1}$. (The figure for other set of ${\rm k}$ one can download from http://astro.franko.lviv.ua/$\sim$novos/fig2.pdf)}
\label{ebg}
\end{figure}

Let us calculate the evolution of number density perturbations of free electrons $\delta_e$ in the region of positive initial matter density perturbation ($\delta_{b}(z_{init},k)>0$, $z_{init}\gg z_{dec}$) of different scales in the $\Lambda{\rm CDM}$-model for the range of redshifts $200\le z \le 10000$. For this we will integrate the system of equations (\ref{SahadxHeIII})-(\ref{eqdTm}) using the code $drecfast.f$.
Results of calculations of number density perturbations of free electrons $\delta_e$, baryon density perturbations $\delta_{b}$ and radiation density perturbations for adiabatic perturbations  with wave numbers ${\rm k}=0.01,...0.1$ Mpc$^{-1}$ are shown in Fig.\ref{ebg}. The visibility function $d\tau /dz e^{-\tau}$ is also shown, its peak denotes the position of last scattering surface.  In $\Lambda{\rm CDM}$-model with parameters \cite{apunevych2007} it is at $z_{dec}=1080$, the wave number corresponding to radius of acoustic horizon at this moment equals ${\rm k}^s_{dec}\simeq 0.037$Mpc$^{-1}$, to particle horizon is ${\rm k}_{dec}\simeq 0.021 $Mpc$^{-1}$.

As one can see in Fig.\ref{xe}, before beginning of recombination of hydrogen ($z\ge 1500$) the amplitudes and phases of relative perturbations of free electrons $\delta_e$ and baryons $\delta_{ b}$ number density coincide for all scales: $\Delta_e\approx 0$. At the epoch of cosmological recombination the relation between them depends on a scale of perturbations. If ${\rm k}\le {\rm k}^s_{dec}$ then amplitude of electron number density relative perturbations  is approximately 4 times higher than amplitude of baryon matter relative perturbations. For smaller scales, ${\rm k}>{\rm k}^s_{dec}$, difference between the values of amplitudes of free electrons and baryons relative perturbations is determined by phase of oscillation of temperature perturbation. After recombination $\delta_e<\delta_{ b}$ for all scales. It is so because we analyse here the adiabatic positive initial perturbations for which the cold dark matter density perturbation increases all the time and after recombination baryon matter falls into the potential well caused by cold dark matter perturbations. So, shortly after recombination the values of baryon density perturbations acquire the same sign and values that CDM ones and practically do not depend on  phase of oscillation at the moment of decoupling.

\begin{figure}
\centerline{\includegraphics[height=6cm]{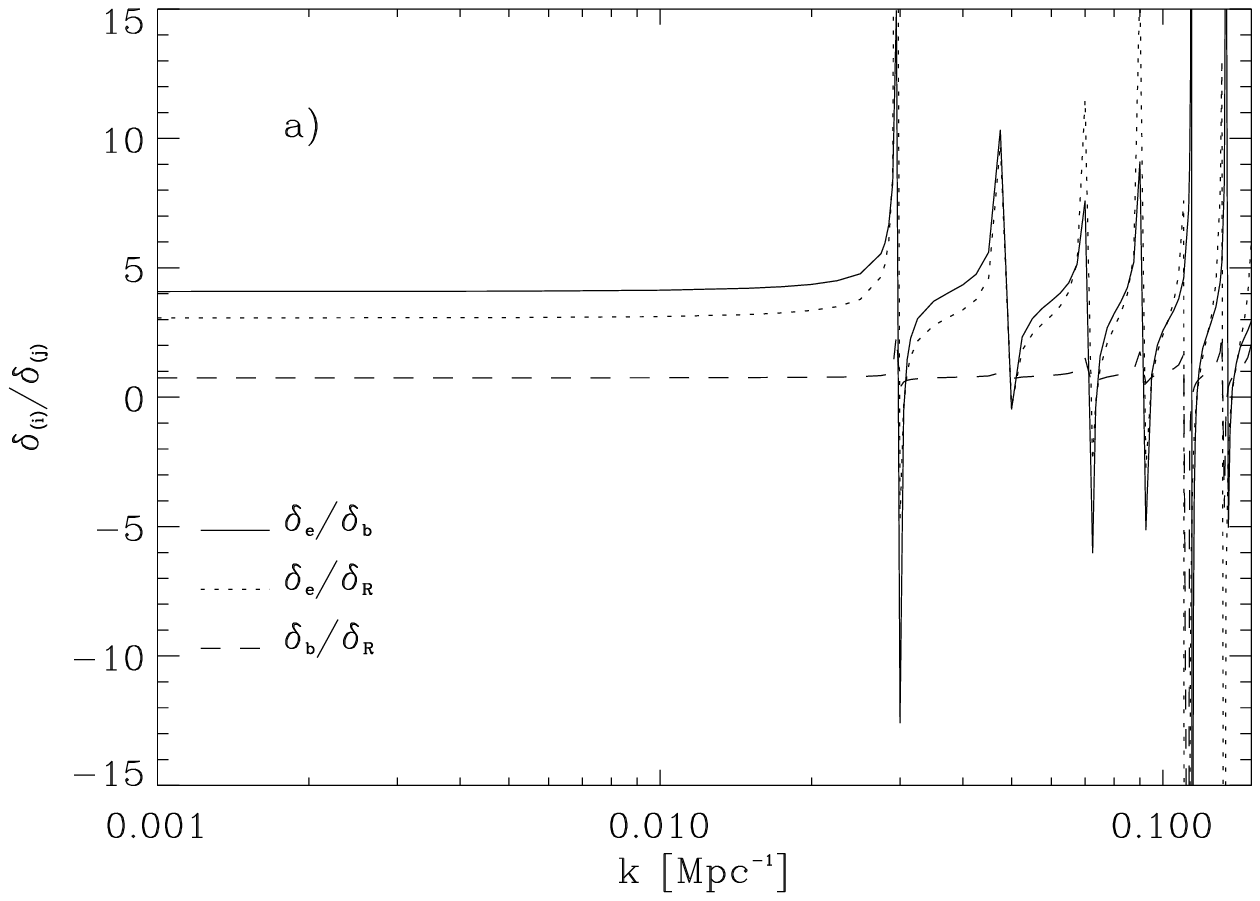}
\includegraphics[height=6cm]{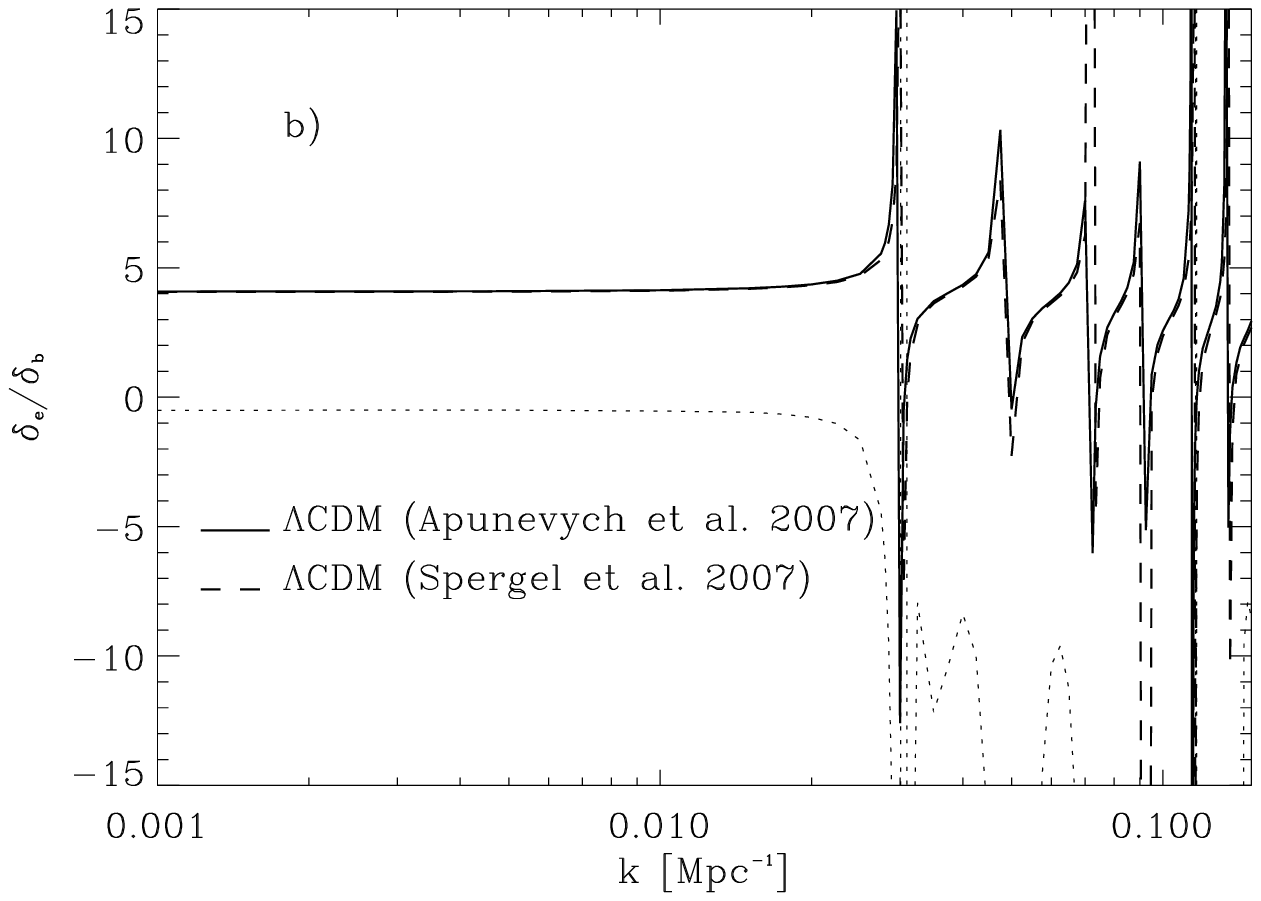}}
\caption{a) The relation of amplitude of free electrons number density perturbations and amplitude of perturbations of total baryon density, $\delta_{e}/\delta_{b}$ (solid line), the relation of amplitude of electron density perturbations and amplitude of radiation perturbations, $\delta_{e}/\delta_{R}$ (dotted line), the relation of amplitude of perturbations of total number density of baryons and amplitude of radiation perturbations, $\delta_{b}/\delta_{R}$ (dashed line). b) The relation of amplitude of free electrons number density perturbations and amplitude of baryon number density perturbations $\delta_{e}/\delta_{b}$ for $\Lambda$CDM-model with parameters $\Omega_{\Lambda}=0.736$, $\Omega_m=0.278$ $\Omega_{b}=0.05$, ${\rm h}=0.68$ \cite{apunevych2007} (solid line) and $\Omega_{\Lambda}=0.76$, $\Omega_m=0.24$ $\Omega_{b}=0.042$, ${\rm h}=0.73$ \cite{spergel2007} (dashed line). The relative difference in percents is shown by dotted line.}
\label{dedb}
\end{figure}

The main part of the CMB photons was scattered by free electrons in the region of maximum of visibility function. Thus, the relation of amplitudes of relative perturbations of free electrons and baryons number density, $\delta_e/\delta_{b}$, at $z_{dec}$ can define some features of CMB anisotropy. That's why a more detailed analysis will be made exactly for this time moment. In Fig.\ref{dedb}a the relation of amplitudes of electron and total baryon number density perturbations ($\delta_e/\delta_{ b}$) is shown by solid line, the relation of amplitudes of electron and radiation number density perturbations ($\delta_{e}/\delta_{R}$) by dotted one and the relation of amplitudes of baryons total number density and  radiation perturbations ($\delta_{b}/\delta_{R}$) -- by dashed one for range of scales $0.001\le {\rm k} \le 0.15$. For scale larger than particle horizon (${\rm k}< {\rm k}_{dec}$), the relations are approximately scale-independent: 
$\delta_e/\delta_{ b}\approx 4.2$, $\delta_e/\delta_{\rm R}\approx 3.2$ and $\delta_{ b}/\delta_{\rm R}\approx 3/4$. At lower scales (${\rm k}> {\rm k}_{dec}$) they change in wide ranges -- this is determined by the different oscillation phases of electron, baryon and photon components perturbations at $z_{dec}$ for different scales. At Fig.\ref{dedb}a the peaks correspond to close to zero values of amplitudes $\delta_{ b}$ and $\delta_{\rm R}$ (for $\delta_{ b}$ zeros are at ${\rm k}\approx 0.0298,\,\,0.0485,\,\,0.0718,\,\,0.0915,\,\,0.114,\,\,0.134$ Mpc$^{-1}$ and for $\delta_{\rm R}$ at ${\rm k}\approx 0.0296,\,\,0.0483,\,\,0.0710,\,\,0.0907,\,\,0.112,\,\,0.132$ Mpc$^{-1}$). Zeros of $\delta_e$ are displaced to lower scales comparing with them: ${\rm k}\approx 0.0305,\,\,0.0501,\,\,0.0737,\,\,0.0941,\,\,0.117,\,\,0.137$  Mpc$^{-1}$.
 
To estimate the dependence of the $\delta_{e}/\delta_{b}$ at recombination epoch on values of cosmological parameters we calculated this ratio for two $\Lambda$CDM-models mentioned in Section 1 (see also Fig.\ref{xe}). The results of calculations are shown in Fig.\ref{dedb}b. One can see, that for super-horizon perturbations the ratio $\delta_e/\delta_{ b}$ practically do not depend on parameters of cosmological model (the difference is less than 0.5\%). For lower scales the difference is significant ($\ge 10\%$) and increases with decreasing of scale. At zero point of $\delta_{b}$ the "jumps`` are caused by their displacement. 
In $\Lambda$CDM-model with parameter \cite{spergel2007} zeros of $\delta_{b}$ at the peak of visibility function correspond to scales ${\rm k}\approx 0.0302,\,\,0.0489,\,\,0.0726,\,\,0.0924,\,\,0.116,\,\,0.135$  Mpc$^{-1}$, and zeros of
$\delta_{\rm R}$ to scales ${\rm k}\approx 0.0300,\,\,0.0488,\,\,0.0718,\,\,0.0917,\,\,0.114,\,\,0.134$ Mpc$^{-1}$. The zeros $\delta_e$ are at ${\rm k}\approx 0.0310,\,\,0.0506,\,\,0.0745,\,\,0.0951,\,\,0.118,\,\,0.139$ Mpc$^{-1}$.

\section{Power spectrum of number density perturbations of free electrons}

\begin{figure}
\centerline{\includegraphics[height=8cm]{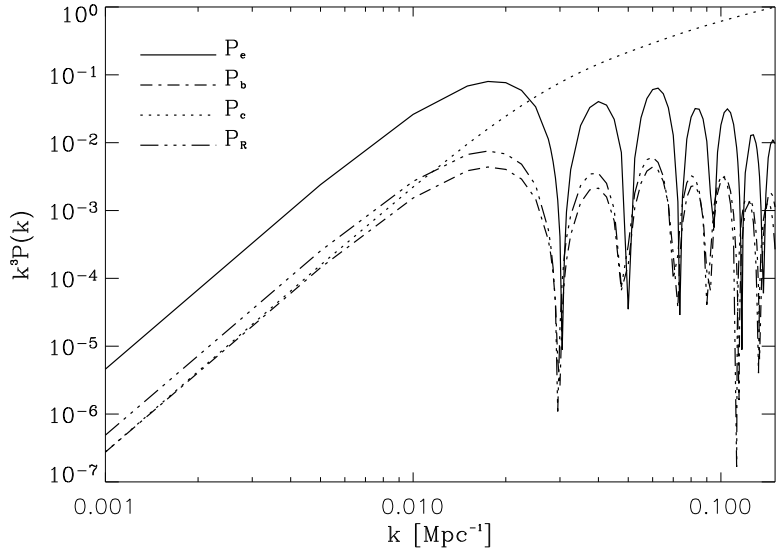}}
\caption{The power spectra of number density perturbations of free electrons (solid line), baryon density perturbations (dash-dotted), perturbations of energy density of thermal radiation (dotted line) at the moment of cosmological recombination $z_{dec}$.}
\label{dke}
\end{figure}

The calculation of the power spectrum of density perturbations of any components at any time moment $t$ 
requires the calculation of the transfer function which is defined as follows:
$$T_i({\rm k},t)\equiv \delta_i({\rm k},t)/\delta_i({\rm k}_{min},t),$$
where ${\rm k}_{min}\ll {\rm k}_{dec}$ and $\delta_i({\rm k},t_{init})=\delta_i({\rm k}_{min},t_{init})$. It means that transfer function is relation of amplitudes of perturbations of two scales ${\rm k}_{min}$ and ${\rm k}$ at any time, initial amplitudes of which were equal at initial time $t_{init}$. If we have the transfer function than the power spectrum of ''i``-th component at $z_{dec}$ can be calculated as follows
$$P_i({\rm k},z_{dec})=A_s{\rm k}^{n_s}T_i^2({\rm k},z_{dec}),$$
where $A_s$ is normalization constant of power spectrum of scalar perturbations, $n_s$ is spectral index. 
Since we analyse here the relation of amplitudes of electron number density and baryon density perturbations, the normalization constant can be  arbitrary (free normalization). The results of calculations of the cosmological perturbations power spectrum of different components are presented in Fig.\ref{dke} by the dimensionless magnitude $P_i({\rm k}){\rm k}^3$ at the moment of cosmological recombination. It summarizes the conclusions
deduced from calculations of evolution of perturbations for different scales (Fig.\ref{ebg}): at moment of cosmological recombination the absolute value of amplitude of free electrons density relative perturbations is few times higher than amplitude of total baryon matter density relative perturbations. It gives also possibility for more detailed analysis of the dependence of the both spectra amplitudes relation on scale of perturbations.
For super-horizon perturbations (${\rm k}\ll {\rm k}_{dec}$) the tilts of spectra are the same for all components $P_i({\rm k})\propto {\rm k}^{n_s}$ and relations of their magnitudes are practically constant: $P_{ b}:P_{\rm c}:P_{\rm R}:P_e\approx 1:1:1.8:17$. At lower scales the power spectrum magnitude of photon-baryon plasma oscillates while one of collisionless component (cold dark matter) increases monotonously. At these scales the relations of power spectrum magnitudes of different components depend on scale of perturbations. The positions of maxima in perturbation power spectra of photon-baryon plasma components approximately coincide:
\begin{itemize}
\item for thermal radiation -- ${\rm k}\approx 0.0175,\,\,0.04,\,\,0.06,\,\,0.08,\,\,0.1,\,\,0.12,\,\,0.1425$ Mpc$^{-1}$;
\item for baryon matter -- ${\rm k}\approx 0.0175,\,\,0.04,\,\,0.06,\,\,0.0825,\,\,0.1025,\,\,0.125,\,\,0.145$ Mpc$^{-1}$;
\item for free electrons -- ${\rm k}\approx 0.0175,\,\,0.04,\,\,0.0625,\,\,0.0825,\,\,0.105,\,\,0.1275,\,\,0.1475$ Mpc$^{-1}$.
\end{itemize} 
For positive cosmological density perturbations ($\delta_c({\rm k},t_{init})=\delta_b({\rm k},t_{init})> 0$) the first maximum is due to the perturbations which were in the phase of first maximal compression at $z_{dec}$; the second one -- to the perturbations which at $z_{dec}$ were in the phase of first maximal decompression; the third one -- to perturbations which at $z_{dec}$ were in the phase of second maximal compression, etc. For adiabatic perturbations with ($\delta_c({\rm k},t_{init})=\delta_b({\rm k},t_{init})< 0$) the maxima have opposite character -- maximal decompression, maximal compression, etc. The positions of ''rifts`` correspond to zeros of perturbations magnitude and coincide with positions of corresponding ``jumps'' in Fig.\ref{dedb}.
Since experimental determination of positions and amplitudes of acoustic peaks of the CMB temperature fluctuations power spectra and their theoretical interpretation are an extremely actual tasks for modern cosmology, we present the ratios of amplitudes of power spectra peaks of different components for two $\Lambda$CDM-models in the Table\ref{peeks}. 
\begin{table}
  \caption{Ratios of peak amplitudes of baryon matter, cold dark matter, thermal radiation and
 free electrons density perturbations power spectra  at $z_{dec}$ for two $\Lambda$CDM-models:  \cite{apunevych2007} -- top line, \cite{spergel2007} -- bottom line.}
  \label{peeks}
  \begin{center}%\footnotesize
    \begin{tabular}{c|c|c}
      \hline %\hline
      & & \\
      N p/p&k [Mpc$^{-1}$]&$P_{ b}:P_{\rm c}:P_{\rm R}:P_e$\\
      & & \\
\hline 
      1&0.0175&1 : \,\,3.7 : 1.70 : 18.2\\
     &0.0175&1 : \,\,3.5 : 1.72 : 18.0\\
\hline 
      2&0.0400&1 : \,67.4 : 1.65 : 18.9\\
   &0.0400&1 : \,69.8 : 1.67 : 18.3\\
\hline
      3&0.0625&1 : \,79.4 : 1.30 : 16.2\\
 &0.0625&1 : \,76.2 : 1.34 : 15.4\\
\hline
      4&0.0825&1 : 203.8 : 1.29 : 13.6\\
 &0.0825&1 : 262.8 : 1.21 : 16.1\\
\hline
      5&0.1050&1 : 231.9 : 0.83 : 11.1\\
 &0.1050&1 : 221.8 : 0.91 : \,10.0\\
\hline
      6&0.1275&1 : 778.7 : 0.63 : 12.2\\
 &0.1275&1 : 685.1 : 0.79 : 10.2\\
      \hline %\hline
    \end{tabular}
  \end{center}
  \vspace{0.3cm}
\end{table}
The comparison of the positions and amplitudes of acoustic peaks of density perturbations power spectra of all components and zeros of $\delta_{\rm R}$, $\delta_{b}$ and $\delta_{e}$ for two models shows that differences do not exceed $\sim 1-2\%$. It explains good coincidence of the predicted power spectra of the CMB temperature fluctuations with observational data: $\chi^2_{min}=37.2$ for $\Lambda$CDM-model with parameters from \cite{apunevych2007} and $37.8$ for $\Lambda$CDM-model with parameters from \cite{spergel2007} for 33 degrees of freedom of the system.

For the first two acoustic peaks the amplitudes of electron number density perturbation power spectrum are $\approx 18$ times higher than amplitudes of perturbations of baryon matter density, for the third one it is $\approx 16$ times higher. For the next peaks such ratios decrease.

Here we do not discuss the relations of amplitudes of baryon, cold dark matter and thermal radiation densities perturbations power spectra because they have been discussed in numerous works (see, for example \cite{hu1995a} and references therein) and are presented here for completeness of the picture.

\section*{Conclusions}

At cosmological recombination epoch the amplitude of relative perturbations of electron number density,  $\delta_e\equiv \delta n_e/n_e$ does not equal the amplitude of relative perturbations of baryon matter density
$\delta_{ b}\equiv \delta n_{b}/n_{ b}$ because the recombination and the photoionization rates have different dependences on density and temperature of baryon-photon plasma. The difference between them becomes prominent when free electrons promptly disappear because of fast recombination of hydrogen at $z\le 1600$ (Fig.\ref{xe} and Fig.\ref{ebg}). At decoupling of thermal radiation from baryon matter the visibility function $d\tau /dz e^{-\tau}$ ($\tau$  is optical depth due to the Thomson scattering by electrons) has maximum at $z_{dec}\approx 1080$ in $\Lambda$CDM-model with parameters \cite{apunevych2007}. The power spectra of relative perturbations of electron number density $P_e({\rm k},z_{dec})\equiv \left\langle \delta_e({\rm k},z_{dec})\delta^*_e({\rm k},z_{dec})\right\rangle$ and baryons $P_{ b}({\rm k},z_{dec})\equiv \left\langle \delta_{b}({\rm k},z_{dec})\delta^*_{b}({\rm k},z_{dec})\right\rangle$ (Fig.\ref{dke}) computed for this moment specify the relation of their amplitudes at different scales.  It is practically flat for perturbations with super-horizon scales (${\rm k}\ll {\rm k}_{dec}$): $P_e({\rm k},z_{dec})/P_{b}({\rm k},z_{dec})\approx 17$. On sub-horizon scales (${\rm k}\ge {\rm k}_{dec}$) the power spectra of electron number density, baryon density and thermal energy density perturbations oscillate. In the $\Lambda$CDM-model the maxima of power spectrum magnitudes of electron number density perturbations at moment of decoupling are at scales ${\rm k}_m\approx 0.0175,\,\,0.04,\,\,0.0625,\,\,0.0825,\,\,0.105,\,\,0.1275,\,\,0.1475$. For them $P_e({\rm k}_m,z_{dec})/P_{b}({\rm k}_m,z_{dec})\approx 18.2,\,\,18.9,\,\,16.2,\,\,13.6,\,\,11.1,\,\,12.2$, correspondingly.  Results and conclusions weakly depend on change of parameters of $\Lambda$CDM-model. 

\section*{Acknowledgments}

This work was supported by the project of Ministry of Education and Science of Ukraine ``The linear and non-linear stages of evolution of the cosmological perturbations in models of the multicomponent Universe with dark energy'' (state registration number 0107U002062) and the research program of National Academy of Sciences of Ukraine ``The exploration of the structure and components of the Universe, hidden mass and dark energy (Cosmomicrophysics)'' (state registration number 0107U007279).

\section*{Appendix} 

 Formulas for calculations of coefficients $A_i,\,\,B_i,\,\,C_i,\,\,D_i,\,\,F_i$:

$$
A_{i} = \left\lbrace \frac{x_{\rm HeII}x_{e}n_{\rm H}\alpha_{\rm HeI}}{x_{\rm HeII}x_{e}n_{\rm H}\alpha_{\rm HeI}-\beta_{\rm HeI}\left(1-x_{\rm HeII} \right) e^{-\frac{h \nu_{\rm HeI2^{1}s}}{kT_m}} };\,\, \frac{x_{\rm HII}x_{e}n_{\rm H}\alpha_{\rm H}}{x_{\rm HII}x_{e}n_{\rm H}\alpha_{\rm H}-\beta_{\rm H}\left(1-x_{\rm HII} \right) e^{-\frac{h \nu_{\rm HI 2s}}{kT_m}} }\right\rbrace ,\nonumber
$$

$$
B_{i} = \left\lbrace \frac{\beta_{\rm HeI}\left(1-x_{\rm HeII} \right) e^{-\frac{h\nu_{\rm HeI2^{1}s}}{kT_m}}}{x_{\rm HeII}x_{e}n_{\rm H}\alpha_{\rm HeI}-\beta_{\rm HeI}\left(1-x_{\rm HeII} \right) e^{-\frac{h \nu_{\rm HeI2^{1}s}}{kT_m}}}; \frac{\beta_{\rm H}\left(1-x_{\rm HII} \right) e^{-\frac{h\nu_{\rm HI 2s}}{kT_m}}}{x_{\rm HII}x_{e}n_{\rm H}\alpha_{\rm H}-\beta_{\rm H}\left(1-x_{\rm HII} \right) e^{-\frac{h \nu_{\rm HI 2s}}{kT_m}}}\right\rbrace ,\nonumber
$$
$$
C_{i} = \left\lbrace \frac{K_{\rm HeI}\Lambda_{\rm He}n_{\rm H}\left( 1-x_{\rm HeII}\right)e^{-\frac{h\nu_{\rm ps}}{kT_m}}}{1+K_{\rm HeI}\Lambda_{\rm He}n_{\rm H}\left( 1-x_{\rm HeII}\right)e^{-\frac{h\nu_{\rm ps}}{kT_m}}};
\frac{K_{\rm H}\Lambda_{\rm H}n_{\rm H}\left( 1-x_{\rm HII}\right)}{1+K_{\rm H}\Lambda_{\rm H}n_{\rm H}\left( 1-x_{\rm HII}\right)}\right\rbrace ,\nonumber
$$
$$
D_{i} = \left\lbrace \frac{K_{\rm HeI}\left( \Lambda_{\rm He}+\beta_{\rm HeI}\right) n_{\rm H}\left( 1-x_{\rm HeII}\right)e^{-\frac{h\nu_{\rm ps}}{kT_m}}}{1+K_{\rm HeI}\left( \Lambda_{\rm He}+ \beta_{\rm HeI}\right) n_{\rm H}\left( 1-x_{\rm HeII}\right)e^{-\frac{h\nu_{\rm ps}}{kT_m}}};\frac{K_{\rm H}\left( \Lambda_{\rm H}+\beta_{\rm H}\right) n_{H}\left( 1-x_{\rm HII}\right)}{1+K_{\rm H}\left( \Lambda_{\rm H}+ \beta_{\rm H}\right) n_{\rm H}\left( 1-x_{\rm HII}\right)}\right\rbrace,\nonumber
$$
where $\nu_{\rm HeI2^{1}s}$ is frequency of HeI $2^{1}s-1^{1}s$ transition, $\nu_{\rm HI 2s}$ is frequency of HI $2s-1s$ transition. For helium in contrary to hydrogen it is needed to take into account the slitting of HeI $2^{1}p$ and $2^{1}s$ that is why there is an additional factor with $\nu_{\rm HeI2^{1}p2^{1}s}=\nu_{\rm HeI2^{1}p}-\nu_{\rm HeI2^{1}s}\equiv \nu_{\rm ps}$ in (\ref{eqHeII}) comparing to (\ref{eqHII}). The values of function $\Theta_i$ are: $\Theta_{\rm  HeII}=1$ and $\Theta_{\rm of HII}=0$. 

Variations of photoionization coefficients were calculated as follows:
$${\delta \beta_i \over \beta_i}= {\delta \alpha_i \over \alpha_i }+{3\over 2}\delta_{T_m}+{h\nu_{2si}\over kT_m}\delta_{T_m},$$
where $h\nu_{2si}$ is ionization energy from 2s state.  
Variations of values of recombination coefficients and matter temperature perturbations are connected by
$$
\frac{\delta \alpha_i}{\alpha_i} = F_{i}\delta_{T_m},
$$
where
$$
F_{i} = \left\lbrace -{1\over 2}\left(1+{(1-p)\sqrt{T_m/T_2}\over 1+\sqrt{T_m/T_2}}+
  {(1+p)\sqrt{T_m/T_1}\over 1+\sqrt{T_{\rm
        m}/T_1}}\right);\,\, \left(b-{d\cdot c\cdot t^d\over 1+c\cdot t^d}\right)\right\rbrace .
$$
\begin{table}[ht]
  \caption{Coefficients and atomic constants of approximate formulas }
  \label{tab}
  \begin{center}%\footnotesize
    \begin{tabular}{c|c|c}
      \hline %\hline
      & & \\
      Constant&Value&Reference\\
      & & \\
      \hline 
      & & \\
      $\chi_{\rm HI}$&$2.17871122\cdot 10^{-18}$ J & \cite{seager1999} \\
      $\chi_{\rm HeI}$&$3.9393393\cdot 10^{-18}$ J & \cite{seager1999} \\
      $\chi_{\rm HeII}$&$8.71869443\cdot 10^{-18}$ J & \cite{seager1999} \\
      $h\nu_{\rm HI 2s}$&$1.63403509\cdot 10^{-18}$ J & \cite{seager1999} \\
      $h\nu_{ps}$&$3.30301387\cdot 10^{-18}$ J & \cite{seager1999} \\
      $h\nu_{\rm HeI2^1s}$&$9.64908312\cdot 10^{-20}$ J & \cite{seager1999} \\
      $h\nu_{2s-1s}$ (HI)&$5.4467613\cdot 10^{-19}$ J & \cite{pequignot1991} \\
      $h\nu_{2s-1s}$ (HeI)&$6.36325429\cdot 10^{-19}$ J & \cite{hummer1998} \\
      $\lambda_{\rm H2p}$&121.567 nm & \cite{seager1999,verner1996} \\
      $\lambda_{\rm HeI2^1p}$&58.4334 nm & \cite{seager1999,verner1996} \\
      F&1.14& \cite{seager1999} \\
      a&4.309& \cite{pequignot1991} \\
      b&-0.6166& \cite{pequignot1991} \\
      c&0.6703& \cite{pequignot1991} \\
      d&0.5300& \cite{pequignot1991} \\
      q&$1.80301774\cdot 10^{-17}$& \cite{hummer1998} \\
      p&0.711& \cite{hummer1998} \\
      $T_1$&$1.30016958\cdot 10^5$ K& \cite{hummer1998} \\
      $T_2$& 3K& \cite{hummer1998} \\
      $\Lambda_{\rm H}$&8.22458 $s^{-1}$&\cite{goldman1989} \\
      $\Lambda_{\rm He}$&51.3 $s^{-1}$&\cite{drake1969} \\
      & & \\
      \hline %\hline
    \end{tabular}
  \end{center}
  \vspace{0.3cm}
\end{table}

\end{document}